\documentclass[twocolumn]{aastex6}
\usepackage{comment}
\usepackage{float}
\usepackage{lineno}

\shorttitle{Drifting subpulses in J1750$-$3503}
\shortauthors{Szary et al.}

\newcommand{\J}{\mbox{J1750$-$3503}}

\begin{document}

\title{MeerKAT observations of the reversing drifting subpulses in PSR \J}
\author{Andrzej Szary\altaffilmark{1, 2}, Joeri van Leeuwen\altaffilmark{2,3}, Geoff Wright\altaffilmark{4}, Patrick Weltevrede\altaffilmark{4}, Crispin H. Agar\altaffilmark{4}, Caterina Tiburzi\altaffilmark{2}, Yogesh Maan\altaffilmark{5, 2}, Michael J. Keith\altaffilmark{4}}
\email{a.szary@ia.uz.zgora.pl}

\altaffiltext{1}{Janusz Gil Institute of Astronomy, University of Zielona G\'ora, Licealna 9, 65-417 Zielona G\'ora, Poland}
\altaffiltext{2}{ASTRON, Netherlands Institute for Radio Astronomy, Oude Hoogeveensedijk 4, 7991 PD, Dwingeloo, The Netherlands}
\altaffiltext{3}{Anton Pannekoek Institute for Astronomy, University of Amsterdam, Science Park 904, 1098 XH Amsterdam, Netherlands}
\altaffiltext{4}{Jodrell Bank Centre for Astrophysics, The University of Manchester, Alan Turing Building, Manchester, M13 9PL, United Kingdom}
\altaffiltext{5}{National Centre for Radio Astrophysics (NCRA - TIFR), Post Bag 3, Ganeshkhind, Pune 411007 India}


\begin{abstract}
We present an analysis of the subpulse drift in PSR J1750$-$3503, 
which is characterized by abrupt transitions of drift direction.
As the pulsar does not exhibit other mode changes or clear nulling, 
it is an ideal candidate system for studying the phenomenon of drift direction change.
For $\sim 80\%$ of the time the subpulses are characterized by positive drift 
-- from early to later longitudes -- while the drift direction is negative in the other $\sim 20\%$.
The subpulse separation for single pulses with positive drift, $P_2=(18.8\pm 0.1)^{\circ}$, is higher then for single pulses with negative drift, \mbox{$P_2=(17.5\pm 0.2)^{\circ}$.}
When the drift is stable, the measured repetition time of the drift pattern is \mbox{$P_3^{\rm obs}=(43.5 \pm 0.4) P$}, where $P$ is pulsar period.
We show that the observed data can be reproduced by a carousel models with subpulse rotation around the magnetic axis using purely dipolar configuration of surface magnetic field.
The observed drift characteristics can be modeled assuming that the actual repetition time $P_3<2P$, such that we observe its aliased value.
A small variation in $P_3$, of the order of $6\%$ (or less assuming higher alias orders), is enough to reproduce the characteristic drift direction changes we observe.
\vspace{0.5cm}
\end{abstract}

\section{Introduction}

\begin{figure*}[!t]
  \centerline{
  \includegraphics[width=\textwidth]{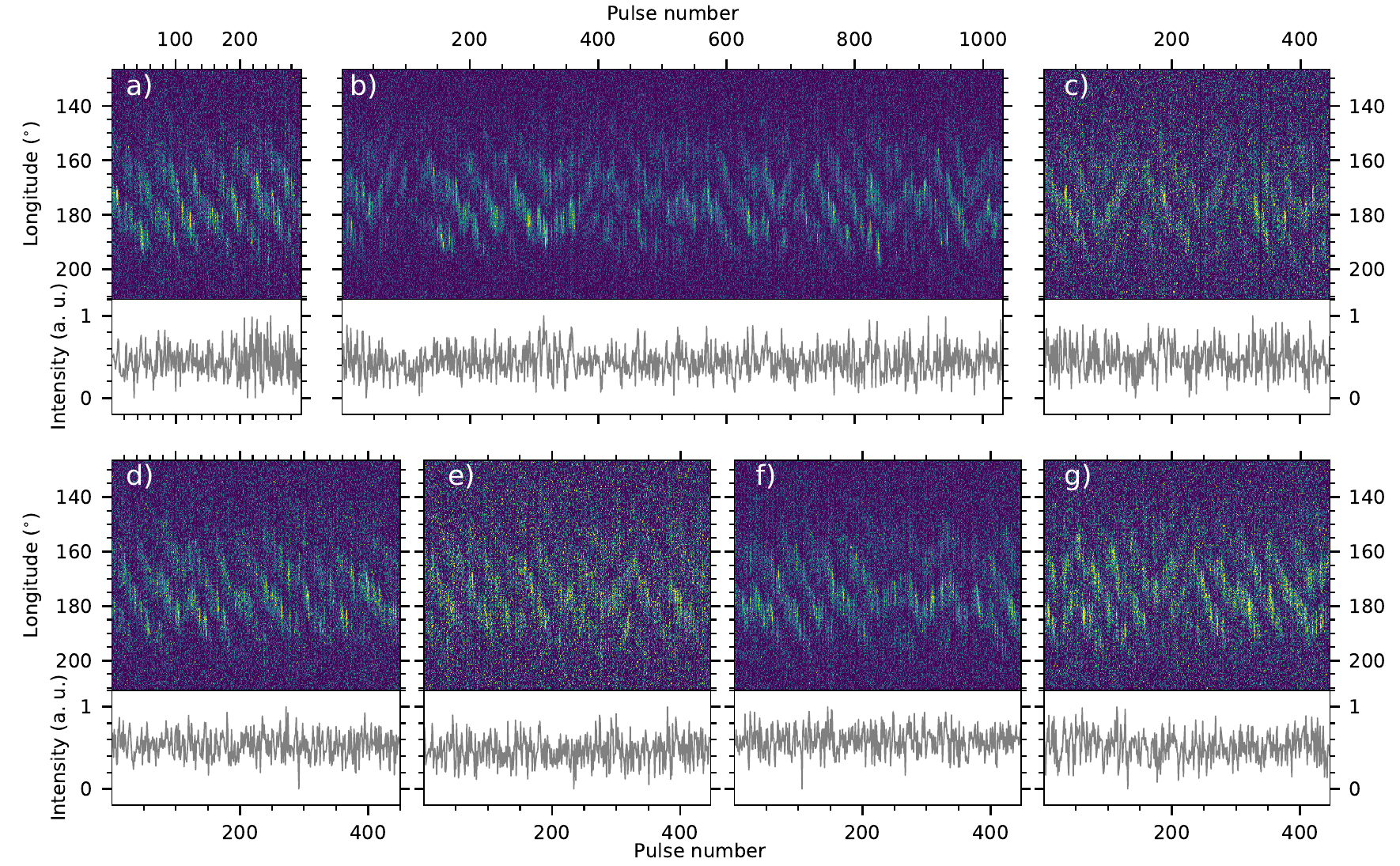}
  }
    \caption{Single pulses in J1750$-$3503 for all observing sessions. Panels from a) to g) correspond to separate observing sessions arranged chronologically (see Table \ref{tab:obs}). For each observing session the main panel shows an intensity plot, while the bottom panel shows variations in the integrated single-pulse intensity.
    }
    \label{fig:single_pulses}
\end{figure*}

The radio emission of pulsars is characterized by a sequence of highly periodic pulses, each consisting of one or more components called \textit{subpulses}.
For a number of pulsars it is known that subpulses appear to ``drift'' across the profile \citep{1968_Drake}.
In the early years of pulsar astronomy, \cite{1975_Ruderman} proposed the carousel model to explain this drifting phenomenon.
That model assumes a pair cascade localized in the form of discharges (called \textit{sparks})
over the polar cap which produce plasma columns along the open magnetic field lines.
The observed radio emission is a consequence of this non-stationary plasma flow and 
this emission is formed 
at altitudes of ${\sim}500$~km above the neutron star surface \citep[see, e.g.,][]{2000_Melikidze, 2003_Kijak, 2017_Mitra}.
Therefore, the location of the sparks at the polar cap determines the location of the subpulses within the pulse profile. 
The carousel model assumes that sparks circulate around the magnetic axis.
Although this assumption  explained a variety of phenomena observed in pulsar data, it lacked the main ingredient: physical justification.
In \cite{2012_Leeuwen} it was noted that the plasma drift velocity relative to the neutron star depends on the variation of the electric potential over the polar cap. Based on this notion, the Modified Carousel (hereafter, MC) model was proposed, where sparks rotate not around the magnetic axis \textit{per se}, but around the location  of the electric potential extremum of the polar cap \citep{2017_Szary}.
If this potential extremum coincides with the center of the polar cap, the subpulses drift simply around the magnetic axis -- even for a non-dipolar surface magnetic field. 

In the recent years, an alternative model gained momentum \citep[see, e.g.][]{2016_Basu, 2020_Mitra, 2020_Basu}.
An important quantity in the drifting phenomenon is the interval between when subpulses 
repeat at the same location in the pulse window, known as the drift periodicity $P_3$.
As the  emission is visible only for a short duration at every pulsar rotation, 
the true value of $P_3$ may be subject to aliasing.
\cite{2016_Basu} postulate the drift is produced as plasma lags behind the corotation of the neutron star, and from the resulting inferred direction of subpulses motion 
the alias order follows.
Given that assumption, \cite{2016_Basu} found a dependence of drift periodicity $P_3$ on the rate of loss of rotational energy $\dot{E}$ (also called the \textit{spin-down luminosity}), $P_3 \propto \dot{E}^{-0.6}$. 
The Lagging Behind Corotation (hereafter, LBC) model poses a challenge to the consensus on the origin of the drifting phenomenon.
It is staggering that more than 50 years of pulsar research has not clarified this key aspect of pulsar radio emission: the true nature of drifting subpulses.
As a part of the Thousand Pulsar Array program \citep{2020_Johnston}, included in the Large Survey Project ``MeerTime'' of the MeerKAT telescope \citep{2020_Bailes}, the low-$\dot{E}$ pulsars have been analyzed to find drifting pulsars that will help to solve this conundrum.

\begin{table*}
\centering \caption{Observation details}
\label{tab:obs}
\begin{tabular}{lcccccc}
\hline
\hline
& & & & & \\
 Obs.ID &  Start time (UTC)  &  Duration  & Number of & Bandwidth & $S/N$ & $N_{\rm ants}$\\
        & (YYYY-MM-DD hh-mm-ss) & (s) & single pulses& (MHz) & &  \\
& & & & & & \\
\hline
& & & & & & \\
20190929-0003 & 2019-09-29 12:41:18 & 202 & 295 & 642 & 110 & 59 \\ 
20191214-0010& 2019-12-14 14:22:12 & 705 & 1031 & 642 & 198 & 59\\ 
20200224-0012& 2020-02-24 03:46:43 & 305 & 446 & 856 & 82 & 29\\ 
20200309-0014& 2020-03-29 04:11:50 & 308 & 450 & 856 & 102 & 28\\ 
20200507-0011& 2020-05-07 23:14:22 & 306 & 447 & 856 & 83 & 28\\ 
20200526-0038& 2020-05-30 22:04:58 & 306 & 447 & 856 & 123 & 31\\ 
20200618-0040& 2020-06-25 21:24:52 & 305 & 446 & 856 & 114 & 30\\ 
& & & & & & \\
\hline
\end{tabular}\\
\end{table*}

For a long time,
the gallery of pulsars with unconventional drifting behavior posed a challenge for theories of the drifting phenomenon.
Some pulsar show variable drift rate (B0031$-$07 \citep{1970_Huguenin, 1997_Vivekanand, 2005_Smits, 2017_McSweeney}, B0809+74 \citep{1983_Lyne, 1984_Davies, 2002_Leeuwen}, B1112+50 \citep{1986_Wright}, J1727$-$2739 \citep{2016_Wen}, J1822$-$2256 \citep{2018_BasuM, 2022_Janagal}, B1918+19 \citep{1987_Hankins, 2013_Rankin}, B1944+17 \citep{1986_Deich, 2010_Kloumann}, B2034+19 \citep{2017_Rankin}, B2303+30 \citep{2005_Redman}, B2319+60 \citep{1981_Wright, 2021_Rahaman}); some display both positive and negative drift directions, simultaneously at different regions of the pulse window, known as the bi-drifting phenomenon (J0815+0939, J1034$-$3224, B1839$-$04, J2006$-$0807 \citep{2005_Champion, 2018_Basu, 2016_Weltevrede, 2019_BasuP}).
Pulsar B0320+39 \citep{2003_Edwards} shows a sudden step of 180 degrees in subpulse phase, and finally pulsars \mbox{B0826$-$34} \citep{1985_Biggs} and \mbox{B0540+23} \citep{1991_Nowakowski} exhibit drift direction changes.
These challenges were addressed using different models or some additional assumptions about conditions around pulsars.
For instance, the bi-drifting phenomenon can be explained using both the LBC and MC models assuming the non-dipolar structure of surface magnetic field \citep[][]{2020_Basu, 2020_Szary}.

In this paper we report an unusual result from our study of \mbox{PSR J1750$-$3503} -- further pulsar showing drift direction changes -- and discuss its importance in unraveling the  nature of the drifting phenomenon.
While a basic drift analysis of \mbox{PSR J1750$-$3503} was part of the overview paper by \cite{2020_Johnston},  
we here present a detailed analysis and interpretation of its drifting behavior.

The paper is organized as follows.
In Section \ref{sec:obs} we show single pulse data of PSR J1750$-$3503 and present the standard analysis, consisting of estimates of the nulling fraction (see Section \ref{sec:nulling}) and pulsar geometry (see Section \ref{sec:geometry}), while in Section \ref{sec:drifting} we show a detailed analysis of drifting subpulses.
The discussion  in Section \ref{sec:discussion} is followed by conclusions in Section \ref{sec:conclusions}.

\section{Observations and data}
\label{sec:obs}


Pulsar J1750$-$3503 was discovered with the Parkes Southern Pulsar Survey \citep{1996_Manchester}.
Its rotational period $P=0.684\,{\rm s}$ and relatively small period derivative, $\dot{P}=3.8\times 10^{-17} \,{\rm s/s }$, place it somewhat close to the graveyard region in the $P-\dot{P}$ diagram.
It is an old pulsar with the characteristic age $\tau_{\rm c}=284 \,{\rm Myr}$, and relatively small rate of rotational energy loss \mbox{$\dot{E}=4.7\times 10^{30}\,{\rm erg\,s^{-1}}$}. 


\subsection{Data taking}

The data used in this paper were recorded with the MeerKAT telescope using the PTUSE\footnote{Pulsar Timing User Supplied Equipment is the SKA1 prototype pulsar
processor developed by Swinburne University of Technology} single-pulse observing mode described in \cite{2020_Bailes}, and were collected during seven observing sessions (see Table \ref{tab:obs}) with the L-band receiver. 
The first two observing sessions were taken using all available MeerKAT antennas, while the following sessions were taken in subarray mode to optimize the telescope use \citep[see][]{2020_Song}.
The central frequency of the observations is 1284\,MHz, while the bandwidth is 642 and 856\,MHz for data collected in 2019 and 2020 respectively and at a $38.3 \mu {\rm s}$ time resolution (see Table \ref{tab:obs} for more details).
The number of frequency channels were 768 and 1024 for the observations in 2019 and 2020, respectively, corresponding to a common channel width of 0.836\,MHz. The data were processed using the MeerTime Single Pulse pipeline (Keith et al., in preparation). The pipeline produces science ready data using 3 stages of processing. In the first stage, the time-series data from the telescope are folded into a single PSRFITS archive using \texttt{dspsr} \citep{2011_Straten}. Polarization calibration is applied, and the frequency channels contaminated by radio frequency interference (RFI) are identified using noise statistics. In the second stage of processing, single-pulse data are formed, after applying the polarization calibration and channel mask obtained from the first stage. The pipeline outputs one set of single-pulse archives that are averaged over the full bandwidth and one set that is frequency-scrunched to 16 sub-bands. The third stage of the pipeline produces a number of diagnostic plots to assess the data quality. More details of this pipeline will be presented elsewhere (Keith et al., in preparation). For the work presented here, we have used the band-averaged single-pulse archives produced by this pipeline.

In Figure \ref{fig:single_pulses} we show sequences of single pulses for all observing sessions.
Except for the first session (the shortest), the collected single pulses are characterized by a very unstable drifting behavior (see panels (b) - (g)).


\subsection{Nulling analysis}
\label{sec:nulling}

\begin{figure}[h!]
  \centerline{\includegraphics[width=\columnwidth]{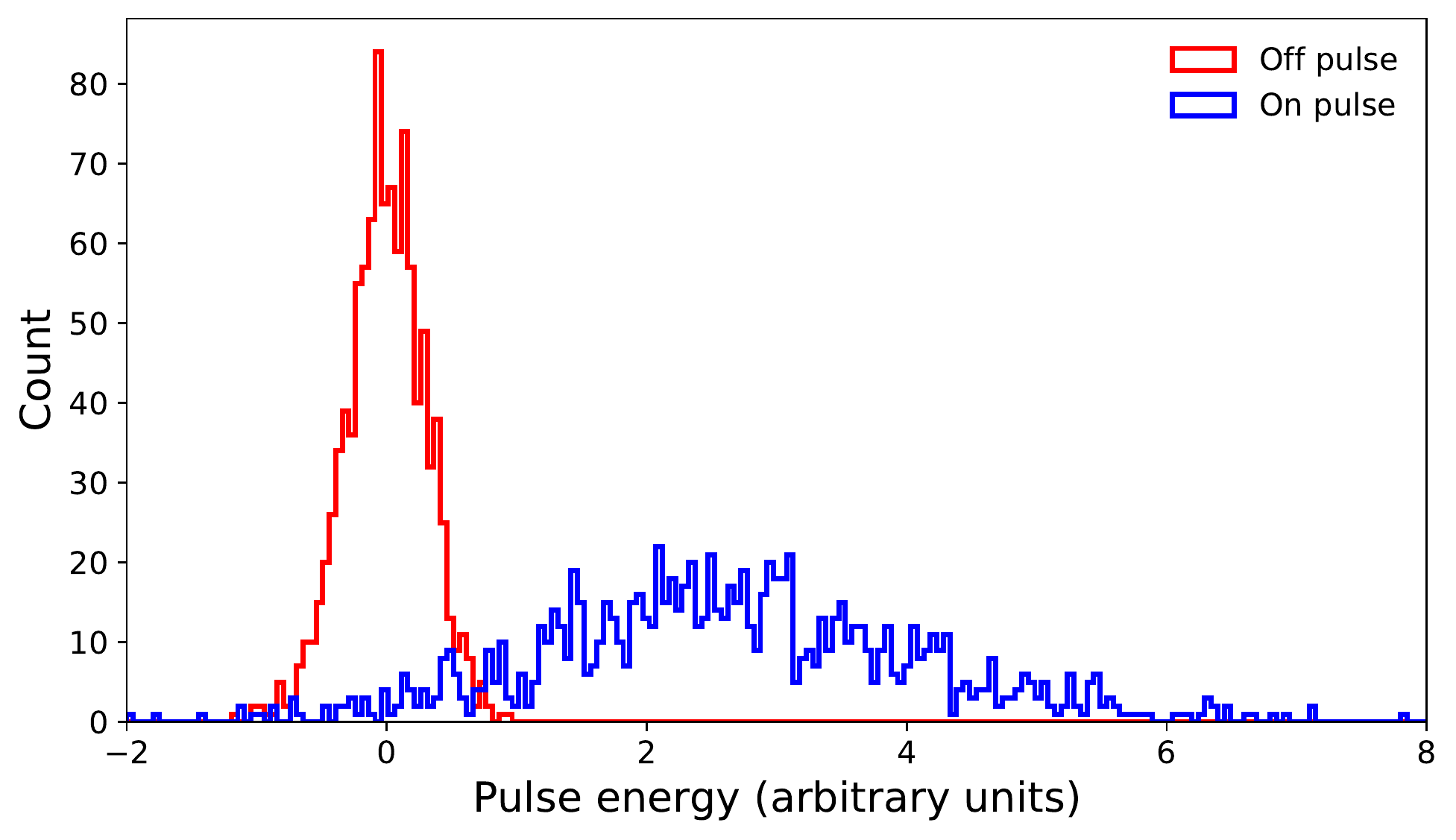}
  }
    \caption{Average energy distribution for on-pulse (red histogram) and off-pulse (blue histogram) regions.
    }
    \label{fig:nulling}
\end{figure}

In Figure \ref{fig:nulling} we show the pulse energy distributions for 1031 pulses of the second observing session as two histograms corresponding to the on and off pulse energies.
The off-pulse distribution is centered around zero and reflects the noise characteristics of the baseline level.
The lack of a bimodal shape in the on-pulse curve indicates that there is no obvious evidence for nulling.


\subsection{Geometry}
\label{sec:geometry}

\begin{figure}[b]
  \centerline{\includegraphics[width=\columnwidth]{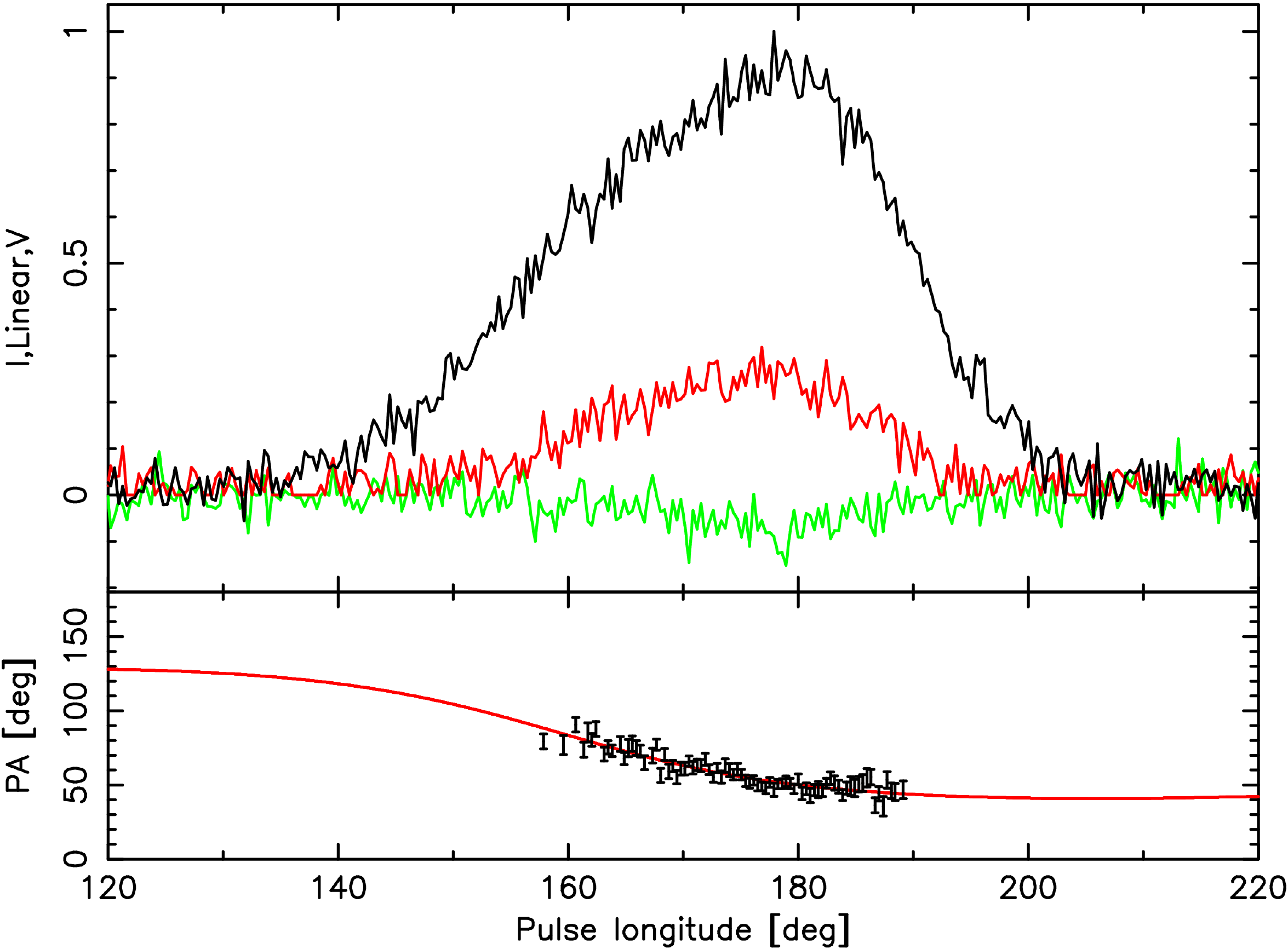}}
    \caption{
    The polarized profile of PSR~J1750$-$3503 (upper panel), and the accompanying PA curve (lower panel). In the profile, the black line shows total intensity whilst the red shows the linear polarization fraction and the green circular. In the lower panel, the black points show the PA as a function of pulse longitude, and the red curve is the best fit of the RVM.
    }
    \label{fig:polarisation}
\end{figure}

Constraining the viewing geometry of PSR~J1750$-$3503 is key to applying the carousel model, and understanding how it gives rise to the observed pattern of drifting subpulses. This geometry is parametrised by two quantities: the magnetic inclination angle $\alpha$, and the impact angle $\beta$ of the observer's line-of-sight. A commonly-used method to determine these is by studying the  pulsar polarization. 

Figure~\ref{fig:polarisation} shows the integrated profile of PSR~J1750$-$3503 and its polarization properties. In the top panel, total intensity is shown by the black curve, the linearly-polarized fraction by the red curve, and the circularly-polarized fraction by the green curve. The lower panel shows how the position angle (PA) of the linear polarization varies as a function of pulse longitude.

There is no evidence of a correlation between the drifting subpulses and polarization properties such as seen in some pulsars when analyzing the individual pulses \citep[e.g. PSR~B0031$-$07;][]{IWJC2020}.

The PA curve was analysed with \mbox{{\sc PSRSALSA}} \footnote{A Suite of ALgorithms for Statistical Analysis of pulsar data. The latest version and a tutorial can be downloaded from https://github.com/weltevrede/psrsalsa}  \citep{2016_Weltevrede} and is shown in Figure~\ref{fig:polarisation}. It is perfectly compatible with the rotating vector model \citep{RCxx1969}, although fitting this was not sufficient to constrain $\alpha$. However, the steepest gradient of the curve was $-2.2\pm 0.3$, which is equal to $\sin\alpha/\sin\beta$ \citep{Kxxx1970}. Better constraints can be placed on $\alpha$ and $\beta$ by taking into account the large observed profile width $W$. The width of the profile at 10~per~cent of its peak intensity is $W_{10} = 58 \pm 7^\circ$, found by fitting a model comprised of two von Mises functions.
The pulse profile width that is observed depends on how the LOS intercepts the conical emission beam, as determined by the viewing geometry $\alpha$ and $\beta$. \citet{GGRx1984} showed that the profile width $W$ can be determined by
\begin{equation}
\label{eq: allowed geometry}
    \cos\rho = \cos\alpha\cos(\alpha+\beta)+\sin\alpha\sin(\alpha+\beta)\cos\bigg(\frac{W}{2}\bigg),
\end{equation}
where $\rho$ is the half-opening angle of the cone of radio emission.
Radio emission of a given frequency is assumed to be produced at some altitude $h_\mathrm{em}$ in the magnetosphere, within the region delimited by the tangents to the last open field lines touching the light cylinder which form a conical beam. The half-opening angle $\rho$ is given by
\begin{equation}
\label{eq: cone angle}
    \rho \approx \sqrt{\frac{9\pi h_\mathrm{em}}{2cP}}
\end{equation}
in the small-angle limit ($h_\mathrm{em} \ll r_\mathrm{LC}$, e.g. \citealt{Rxxx1990}), where $c$ is the speed of light.
The emission height of radio pulsars is believed to lie within 200 to 400~km \citep{JKxx2019} irrespective of period. This means that for PSR~J1750$-$3503, where $P=0.684$~s, the cone opening angle lies in the range $5^\circ \lesssim \rho \lesssim 10^\circ$ if the emission height is not atypical. 

Combining Equations (\ref{eq: allowed geometry}) and (\ref{eq: cone angle}) leads to a set of constraints such that only certain combinations of $\alpha$ and $\beta$ are ``allowed''. These allowed geometries indicate that $|\beta| < 10^\circ$, and $5^\circ \lesssim \alpha \lesssim 25^\circ$. 
These values are input for the subpulse-drift interpretation and modeling that we pursue next. 

\begin{figure*}[!t]
  \centerline{
  \includegraphics[width=\textwidth]{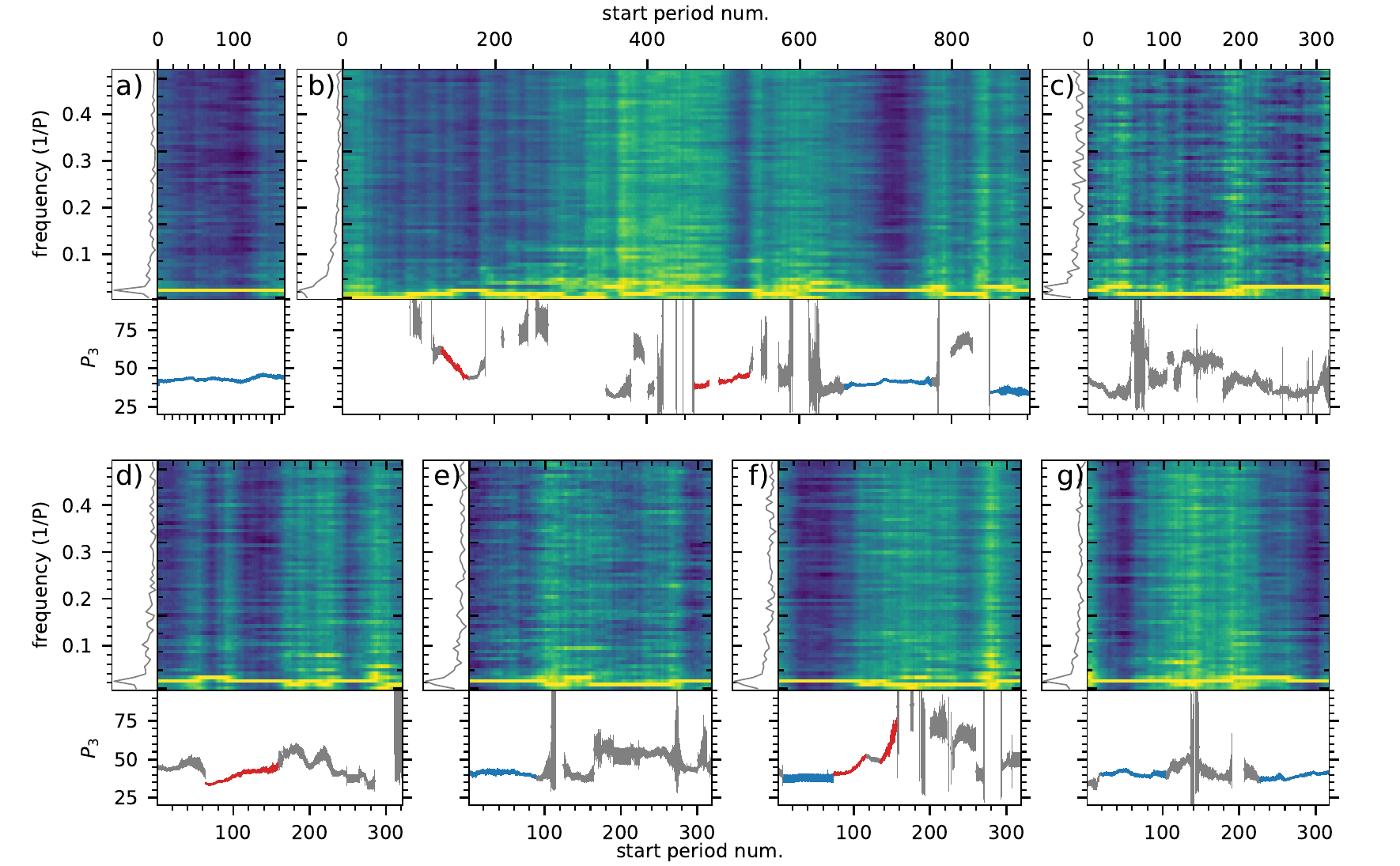}
  }
    \caption{Variation of the fluctuation spectra as a function of the start period (the number of the first pulse in the interval) for all observing sessions. Panels from a) to g) correspond to single pulses shown in Figure \ref{fig:single_pulses}. The fluctuation spectra are determined for 128 consecutive single pulses.
    For a single observing session the left panel shows the time average fluctuation spectrum, while the measured $P_3$ values are shown in the bottom panel. The blue and red points in the bottom panels correspond to sections with stable $P_3$ and sections with monotonically increasing/decreasing $P_3$, respectively. The uncertainties correspond to one sigma.
    }
    \label{fig:p3_evolution}
\end{figure*}


\begin{figure*}[!t]
  \centerline{
  \includegraphics[width=\textwidth]{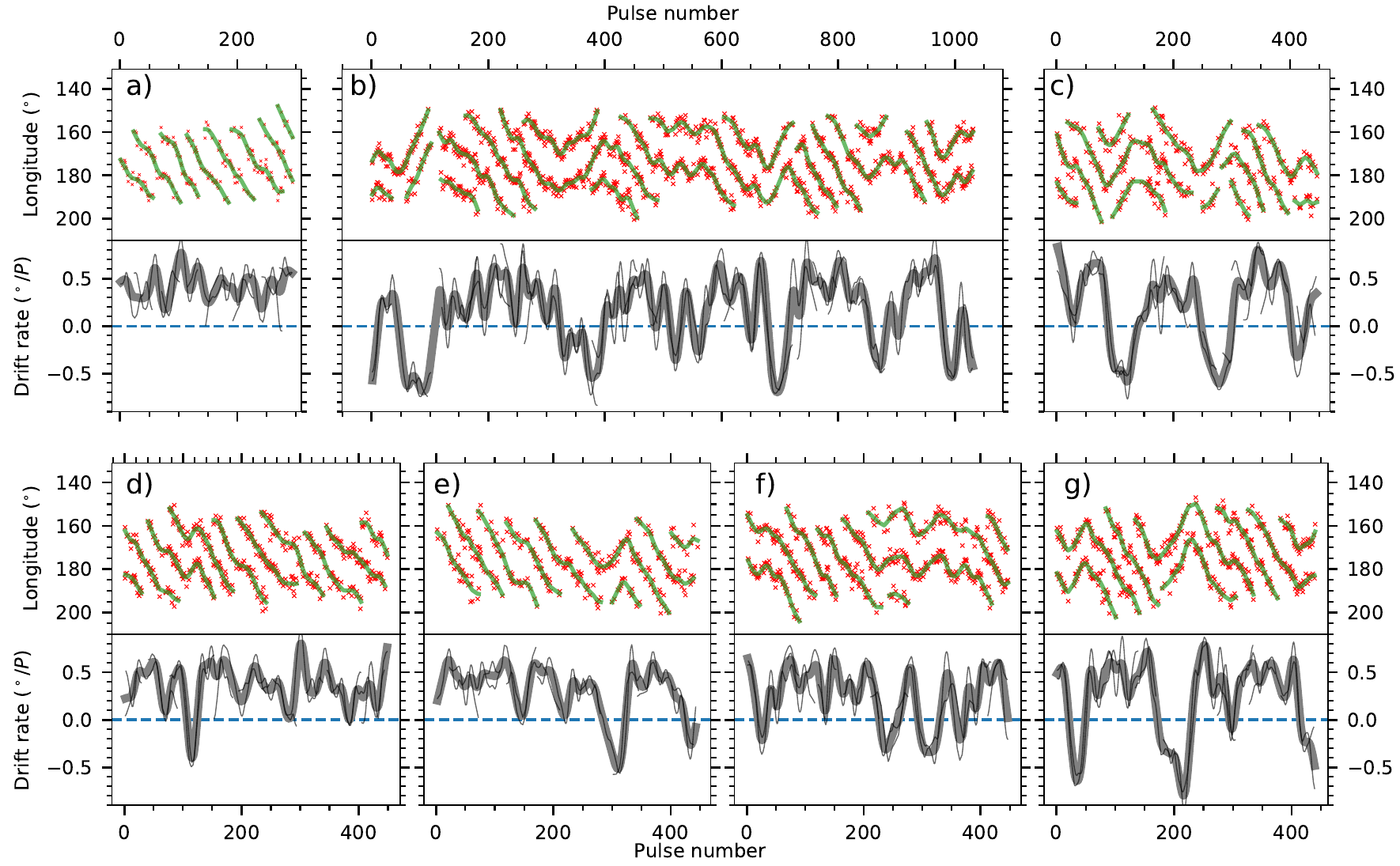}
  }
    \caption{
    Drifting patterns (the top panels) and drift rates (the bottom panels) for single pulses shown in Figure \ref{fig:single_pulses}. 
    The red crosses show location of subpulses, while the green lines correspond to fits to driftbands (see text for more details). 
    The black solid lines in the bottom panels show drift rates of an individual track, while the solid gray line shows the average drift rate.
    }
    \label{fig:driftrates}
\end{figure*}

\newpage

\section{Subpulse Drifting}
\label{sec:drifting}

In this section we present a detailed analysis of the drifting subpulses.

\subsection{Drift characteristics}

To study the drift characteristics in \J~ we use  Longitude Resolved Fluctuation Spectra \citep[LRFS,][]{1970_Backer}, whose computation involves discrete Fourier transforms of consecutive pulses along each rotational-phase longitude.
To find the time intervals where the  drifting behavior is stable, we calculate the LRFS of intervals of 128 single pulses, shifting the starting point by one period every time \citep[see also][]{2009_Serylak}.
The repetition time of the drift pattern, $P_3$, is determined by fitting a Gaussian near the major peak in the longitude averaged fluctuation spectra. 
In Figure \ref{fig:p3_evolution} we show the variation of the fluctuation spectra as a function 
of the pulse number corresponding to the interval start.
Visual inspection of the drift in Figure \ref{fig:single_pulses}
shows the  patterns are not always stable, but vary.
The continuous range of these periodicities is visible in the bottom panels of 
Figure \ref{fig:p3_evolution}. 
Below we focus on a number of notable periodicity epochs within each session.
The repetition time seems to be stable only for the first observing session, with $P_3=(43\pm 1) P$.  
To calculate the repetition time for the first session with a better accuracy we use the \mbox{PSRSALSA} software package \citep[see][for more details]{2016_Weltevrede}, which results in $P_3=(43.5 \pm 0.4) P$.
The measurements for the other observing sessions are characterized by much more variability, resulting in 
session-averaged repetition times
$P_3=(47\pm 14) P$, $P_3=(43\pm 8) P$, $P_3=(44\pm 6) P$, $P_3=(47\pm 7) P$, $P_3=(50\pm 13) P$, and $P_3=(40\pm 3) P$ respectively for the observing sessions presented in panels from b) to g).
The errors on these mean values are the one sigma standard deviation. They signify the variation in the underlying pattern.
During those sessions, a few intervals with a stable $P_3$ can be identified (see Figure \ref{fig:p3_evolution}). 
These show steady repetition values of $P_3=(43 \pm 1) P$, $P_3=(41 \pm 1) P$, $P_3= (35 \pm 1) P$, $P_3= (41 \pm 1) P$, $P_3= (37.8 \pm 0.3) P$, $P_3= (41 \pm 1) P$, and $P_3= (39 \pm 2) P$, respectively, for the starting pulse numbers 660-773 and 852-903 in panel b), 1-88 in panel e), 8-73 in panel f), and 20-103 and 228-318 in panel g).
Furthermore, a number of monotonically increasing and decreasing $P_3$ values are identified. 
In panel b), between pulse numbers 131-165 the repetition time $P_3$ decreases from $(61 \pm 3) P$ to $(44 \pm 1) P$, and between 463-534 $P_3$ increases from $(39 \pm 2)$ to $(46 \pm 2)P$. Between the starting pulse 64-158 in panel d) $P_3$ increases from $(34 \pm 1) P$ to $(46 \pm 4)P$. In panel f), between pulses 74-115 $P_3$ increases from $(41 \pm 2)P$ to $(52 \pm 1)P$, and between pulses 135-156 increases from $(49 \pm 2)P$ to $(72 \pm 5) P$.

We identify seven sections with relatively stable $P_3$ and five sections with  monotonically increasing/decreasing $P_3$, while no evidence of periodic changes in the measured $P_3$ values was found.
Only during the first observing session the duration when $P_3$ is stable is larger than the length of LRFS.
Except the first observing session, inspection of single pulses (see Figure \ref{fig:single_pulses}) shows that time intervals with both stable and changing $P_3$ are characterized with drift direction changes.
Since the analysis performed using the LRFS is limited by the length of the Fourier transformation, we perform the drift analysis using single-pulses in the following section. 

\newpage

\subsection{Single-pulse analysis}
\label{sec:driftrates}

\begin{figure}[tbh]
  \centerline{\includegraphics[width=\columnwidth]{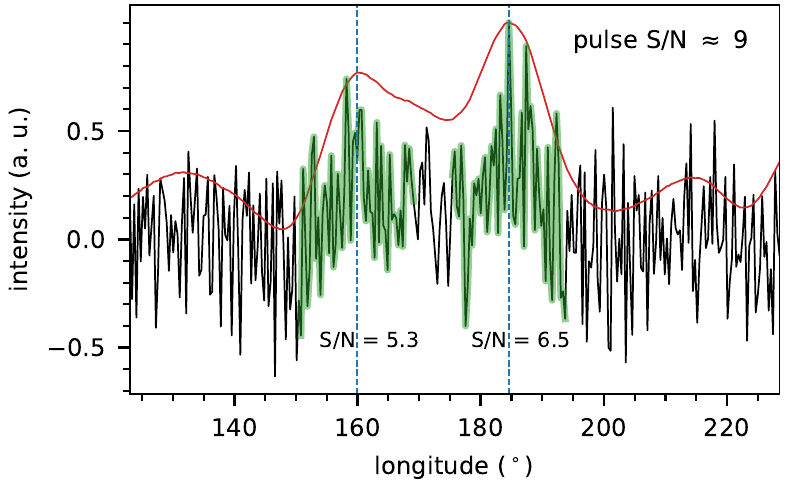}}
    \caption{Subpulse detection procedure. The black solid line shows a single pulse, while the red solid line corresponds to the convolution of the single pulse with a Gaussian function. The green lines mark the subpulse widths, while the blue dashed lines correspond to subpulse locations.
    }
    \label{fig:subpulse_detection}
\end{figure}

To find the subpulse location (see the red crosses in the upper panels of Figure \ref{fig:driftrates}), the data of a single pulse are convolved with a Gaussian function having the mean width of the subpulses.
The peaks in the convolution are used to determine the longitudes of the subpulses.
In the analysis presented in this paper we consider only subpulses with $S/N$ higher than five.
Furthermore, the detected peaks are visually inspected and only subpulses that are located together as a drift band, and are within the expected longitude range, are considered in further analysis. 
In Figure \ref{fig:subpulse_detection} we show the detection procedure for a sample single pulse, which results in the following detection fractions: $70\%$, $83\%$, $69\%$,  $74\%$, $69\%$, $76\%$, $79\%$, for all observing sessions, i.e. data presented in panels (a) to (g), respectively, in Figures \ref{fig:single_pulses} and \ref{fig:driftrates}. 

\newpage
\subsubsection{Drift rate}

The drift rate is defined as \mbox{$D = \Delta \phi$} per pulse period ($^\circ / P$), where $\Delta \phi$ is the longitude shift in degrees during one pulse period $P$.
A positive value indicates a drift from early to later longitudes, while a negative value corresponds to a drift from late to earlier longitudes.
To calculate the drift rates we apply a cubic smoothing spline estimate for every drift track using the {\it SmoothingSplines}\footnote{Package for nonparametric regression with Cubic Smoothing Splines. The latest version can be downloaded from https://github.com/nignatiadis/SmoothingSplines.jl} software package and calculate its gradient. 
The calculated value depends on the smoothing parameter, $\lambda$.
In Figure \ref{fig:driftrates} we show the drift tracks (green lines in upper panels) and drift rates (lower panels) for all observing sessions, obtained using the smoothing parameter $\lambda=200$.
In total, we find 24 instances with negative drift rate (i.e. subpulses drifting from late to earlier longitudes), with the longest instance lasting for 75 pulses (see pulse number 323 in Figure \ref{fig:driftrates}(b)). 
The positive drift rate occurs more often covering $\sim$78\% of all the observing time, with the longest instance lasting for 295 pulses (see Figure \ref{fig:driftrates}(a)). 
The mean duration of time intervals with negative drift rate, $(28 \pm 4)P$, is considerably smaller than the mean duration of time intervals with positive drift rate, $(88 \pm 15)P$. 
Furthermore, the absolute value of the mean of positive drift rates $|D_{+}| = (0.388 \pm 0.003)^\circ/P$ is higher than the absolute value of the mean of negative drift rates $|D_{-}| = (0.314 \pm 0.006) ^\circ/P$.
There is no evidence of periodicity in the drift rate changes.

\subsubsection{Subpulses separation $P_2$}
\label{sec:p2_single-pulse}

\begin{figure}[t]
  \centerline{
  \includegraphics[width=\columnwidth]{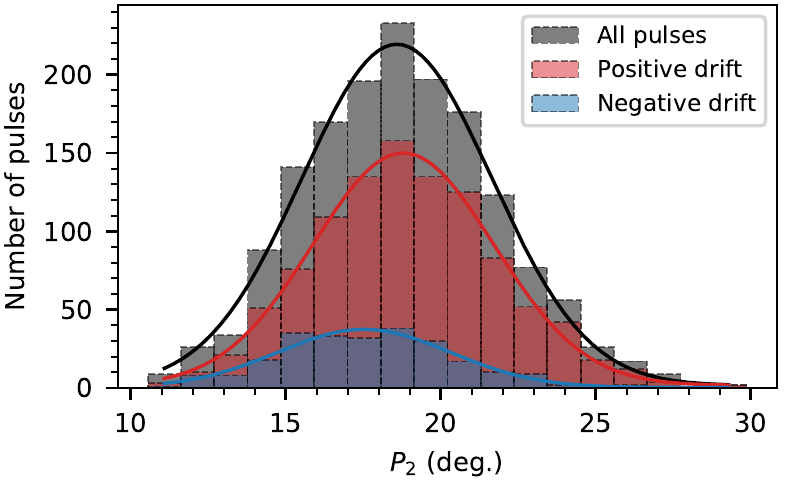}
  }
    \caption{Subpulse separation, $P_2$, for all single pulses (the gray histogram), single pulses with positive (the red histogram) and negative (the blue histogram) drift rate. The black, red and blue solid lines correspond to Gaussian fits for all pulses, positive and negative drift, respectively.
    }
    \label{fig:p2_analysis}
\end{figure}

Another quantity that we can use to characterize the drifting subpulses is the longitudinal separation between subpulses, $P_2$,  or in other words, the separation between adjacent drift bands. 
We use the drift rate information and subpulses locations in single pulse data to analyze the subpulses separation.
Figure \ref{fig:p2_analysis} shows distribution of $P_2$ measurements for all single pulses, $P_2 = (18.58 \pm 0.07)^{\circ}$, single pulses with positive drift, $P_2 = (18.79 \pm 0.07)^{\circ}$, and negative drift $P_2 = (17.52 \pm 0.16)^{\circ}$. The uncertainties correspond to the standard error.


\subsection{Average pulse profiles}
\label{sec:average}

\begin{figure}[h!]
 \centerline{
 \includegraphics[width=\columnwidth]{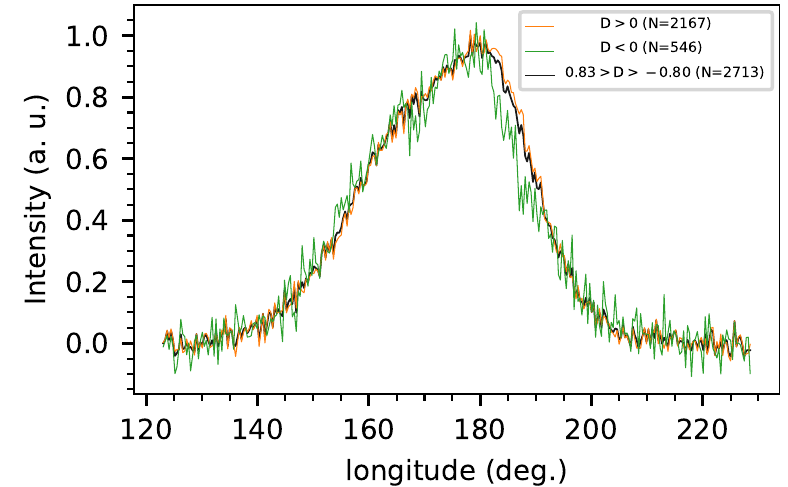}
  }
    \caption{Average profiles produced by single pulses with positive (the orange line) and negative (the green line) drift rates. The black line corresponds to  the average profile for all single pulses.  
    }
    \label{fig:averages}
\end{figure}

Previous studies have shown that for a number of pulsars a variation of drift rate is associated with the changing average profile shape \citep[see, e.g.][and references therein]{2005_Esamdin}.
To show the variation of the average profile in different drift states, in Figure \ref{fig:averages} we compare the average profiles produced by single pulses with positive and negative drift rates.
The profiles were normalized using the number of single pulses used.
The black solid line in the figure corresponds to the average profile for all single pulses.
The average profile for positive drift is characterized with one skewed peak formed from the underlying two peak structure.
On the other hand, in the average profile for single pulses with negative drift the two peak structure is more visible.
Moreover, the maximum peak seems to be shifted towards earlier longitudes.
To estimate differences in intensity we calculate area under the curves, and normalize them assuming that the area under the average profile for all single pulses is equal to one.
The area for positive and negative drift, respectively, is $1.01 \pm 0.06$ and $0.98 \pm 0.06$, here the uncertainties correspond to the standard error.
The area measurements are consistent within the uncertainties, suggesting no significant changes in subpulse brightness during reversals.

\newpage
\subsubsection{Drift direction change}
\label{sec:driftdirection_longitudes}

\begin{figure*}[t]
  \centerline{\includegraphics[width=16cm]{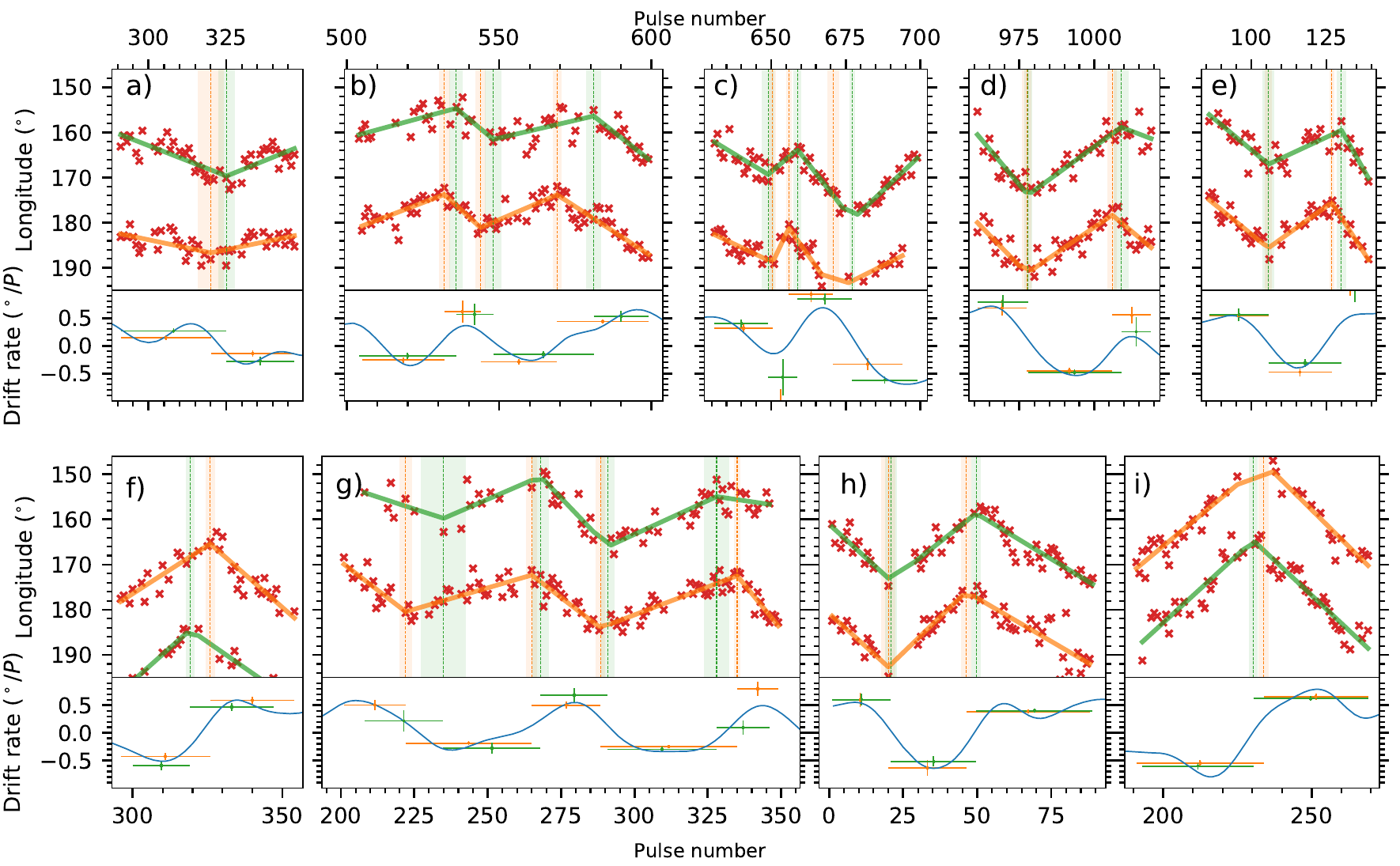}} 
    \caption{Drifting patterns (the top panels) and drift rates (the bottom panels) for time intervals with drift direction changes. 
    The panels correspond to the following observing sessions: (a) - (d) - second ; (e) - fourth; (f) - fifth; (g) - sixth; and (h), (i) - seventh. 
    The red crosses show location of subpulses, while the green and orange lines correspond to fits to driftbands (see text for more details). 
    The blue solid lines in the bottom panels show the average drift rate.
  }
    \label{fig:driftdirection}
\end{figure*}

To explore the nature of drift direction change in more details we use a method based on a linear regression, {\it Segmented}\footnote{Segmented is an R package to fit regression models with broken-line relationships}.
Figure \ref{fig:driftdirection} shows a detailed analysis of subpulses positions for selected time intervals.
We use the coefficient of determination, $R^2$, to check how well the data are replicated by the models.
For all tracks shown in Figure \ref{fig:driftdirection}, $R^2$ is $0.77 \pm 0.11$ and $0.75 \pm 0.14$ for the cubic smoothing spline and linear regression fits, respectively.
The color solid lines in the upper panels of Figure \ref{fig:driftdirection} show the linear regression fits, while the bottom panels show a comparison of drift rates derived by the cubic smoothing spline fits (the solid blue line) and the linear regression fits (the points with error bars).  
The results suggest that the drift in \J~ can be equally well described by either smooth or rapid changes in drift direction.
The vertical dashed lines in the upper panels of Figure \ref{fig:driftdirection} mark the estimated time of drift direction change. 
The breakpoints are consistent within uncertainties throughout all subpulse tracks (see, e.g., pulse number 320 in panel (a), pulse number 975 and 1010 in panel (d), pulse number 21 and 50 in panel (h), of Figure \ref{fig:driftdirection}).


\subsection{Aliasing effect}
\label{sec:aliasing}

The pulsed emission is visible for a short duration every rotation, which could mean that the measured apparent drift rate of drifting subpulses is the result of the aliasing effect.
The shortest actual $P_3$ unaffected by aliasing, i.e. the time in which the consecutive subpulse shifts to the space occupied by the previous one, is twice the pulsar rotation period $P$.
The actual modulation period, $P_3$, can be expressed as $P / P_3=n + P / P_3^{\rm obs}$. 
Here $P_3^{\rm obs}$ is the apparent (possibly aliased) modulation period, where $-0.5 < P/P_3^{\rm obs} < 0.5$. 
The sign of $P_3^{\rm obs}$ changes depending on the apparent sign of the gradient of the drift bands.
The alias order $n$ is the nearest whole number of undetected subpulses from one pulse to the next (see, e.g., \citealt{2002_Edwards}).
Note that the definition of $n$ differs from the alias order introduced in \cite{2004_Gupta}.
In order to estimate the true value of $P_3$ the alias order has to be resolved; but in general it is not always possible to infer the value of $n$ from the observations.
In this section we explore the possibility that the peculiar drift characteristics of J1750$-$3503 are caused by aliasing.

\begin{figure}[bt]
  \centerline{\includegraphics[width=7.3cm]{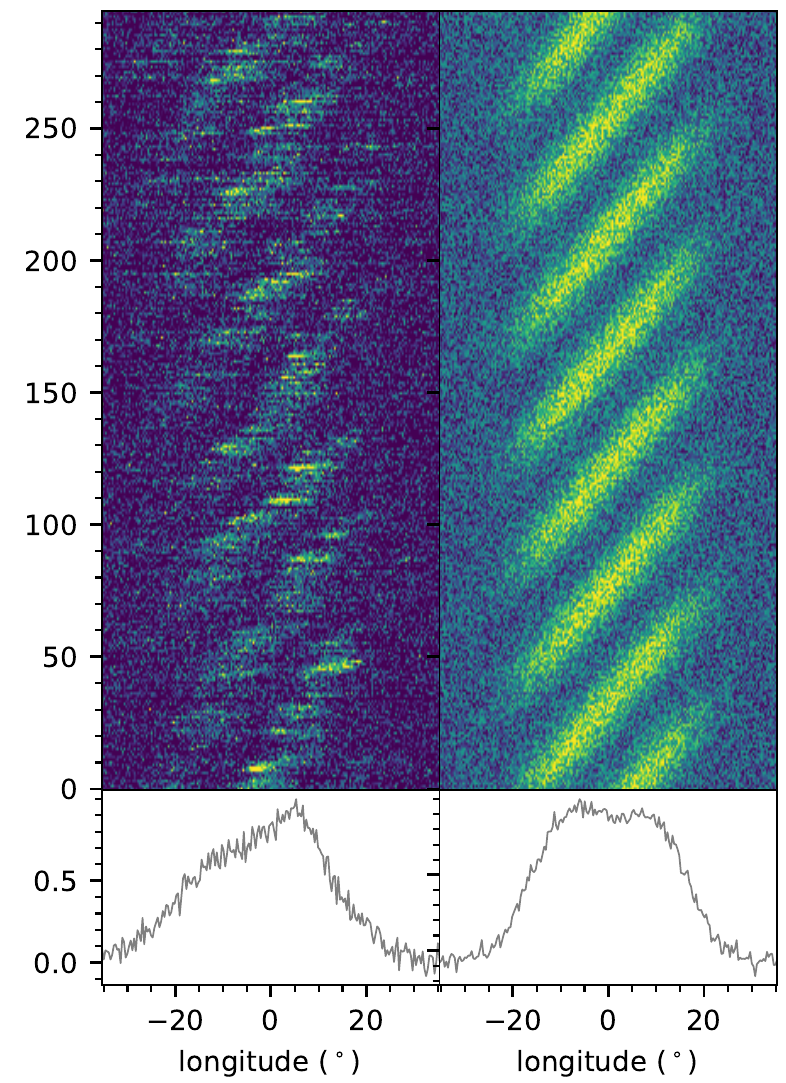}}
    \caption{Single pulses for the first observing session (the left panel) and the modeled single pulses (the right panel). 
    The bottom panel shows the average profile.
    }
    \label{fig:singles}
\end{figure}

To model single pulses we use the approach proposed in \cite{2020_Szary} with solid-body-like rotation of sparks around the center of the polar cap.
In order to achieve the solid-body-like rotation we assumed that the angular velocity of all sparks is equal. 
At times when the drift is relatively stable the measured repetition time $P_3^{\rm obs}\sim 44 P$ (see Figure \ref{fig:p3_evolution} a)).
In Figure \ref{fig:singles} we compare single pulses observed during the first observing session (the left panel) and the modeled single pulses (the right panel).
The pulses are modeled using the following pulsar geometry: inclination angle $\alpha=10^{\circ}$, impact parameter $\beta=-4.5^{\circ}$ and the emission height $h_{\rm em} = 300 \,{\rm km}$, which results in the opening angle $\rho=8.3^{\circ}$ (see Equations \ref{eq: allowed geometry} and \ref{eq: cone angle}) and the measured pulse width $W_{10}=44 \pm 2^{\circ}$. 
To determine the number of subbeams in the carousel, and therefore the sparks at the polar cap, for this specific geometry, we compare phase information obtained using the LRFS of the observed and modeled signals. 
The single pulses are modeled using $P_3=0.9778 P$ resulting in the observed repetition time $P_3^{\rm obs} = (43 \pm 3)  P$. 
We use thirteen subbeams with an imprint on the polar cap with the Gaussian variance of $9\, {\rm m}$ (taking the neutron star radius to be 10 km). 
Note, that since the drifting phenomenon is affected by the viewing geometry \citep{2004_Gupta} the derived number of subbeams is a unique solution only for a given pulsar geometry.

\begin{figure*}[bt]
  \centerline{\includegraphics[width=\textwidth]{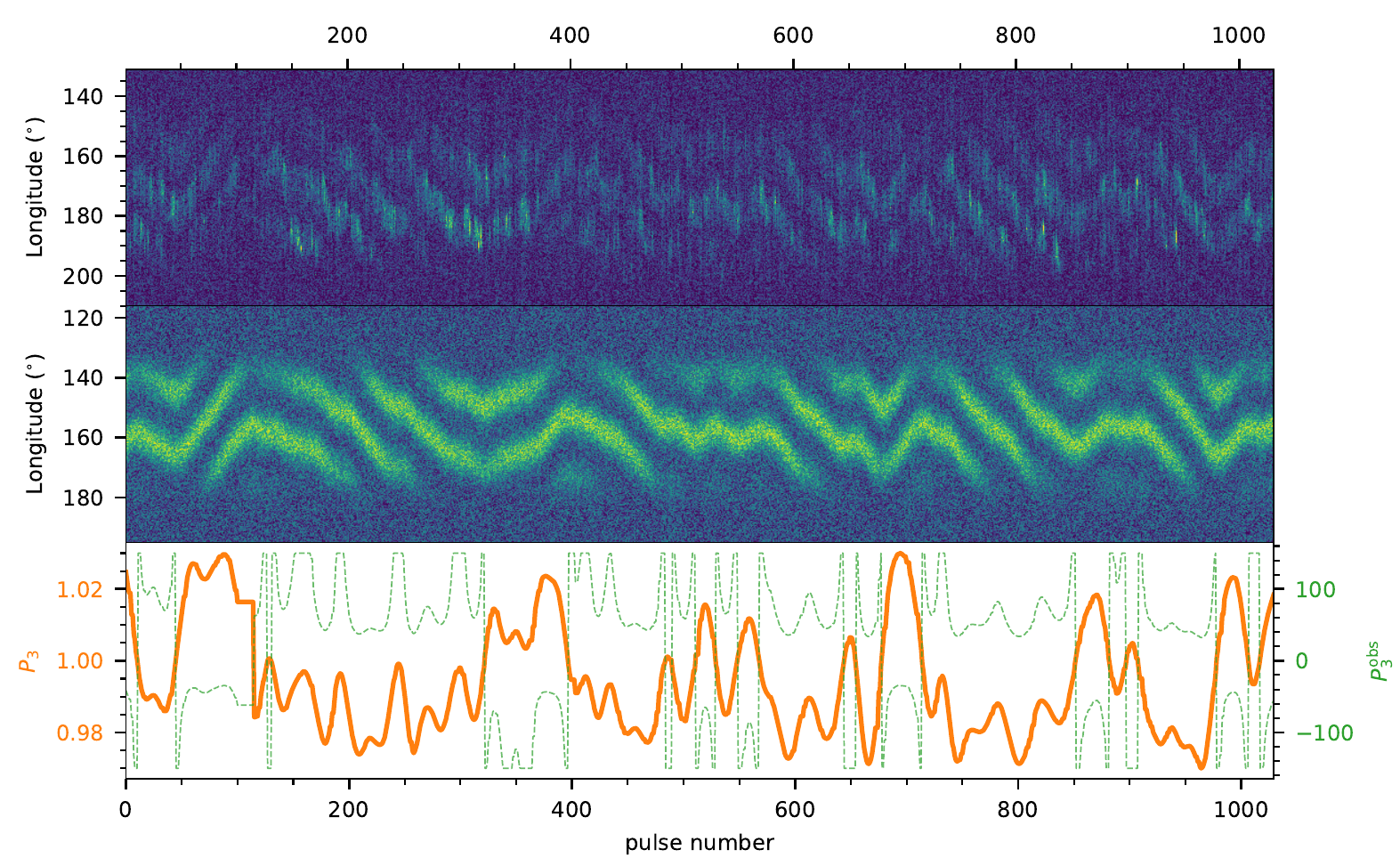}}
    \caption{Single pulses in J1750$-$3503 for the second observing session (the top panel), and the modeled single pulses (the middle panel). 
    The bottom panel shows the assumed repetition time $P_3$ (the orange solid line) and the resulting repetition time $P_3^{\rm obs}$ (the green dashed line).
    }
    \label{fig:obs_model_modeledpulses}
\end{figure*}

At times when the drift in J1750$-$3503 is characterized by non-stable behavior (see, for instance, panel (b) in Figure \ref{fig:p3_evolution}), $P_3^{\rm obs}$ ranges from $\sim 33 P$ to $\sim 85 P$.
Assuming the modulation is observed in the first alias order, and assuming the actual drift direction is the same as that observed ($n=1$) the actual repetition time, $P_3$, ranges from $0.97 P$ to $0.988 P$.
If, on the other hand, we assume that the actual drift direction is opposite to that observed ($n=-1$), the actual repetition time, $P_3$, ranges from $1.011 P$ to $1.031 P$.
As follows from the above calculations, a change in $P_3$ of only about $6 \%$ could produce both the observed drift direction change and the measured $P_3^{\rm obs}$ range.
Furthermore, for higher alias orders an even smaller change in $P_3$ results in the observed drift characteristics.
To model the changing $P^{\rm obs}_3$ we alter the angular velocity of sparks according to the measured drift rates (see the bottom panels in Figure \ref{fig:driftrates}).
In Figure \ref{fig:obs_model_modeledpulses} we compare the single pulses observed during the second observing session (the top panel) with the modeled single pulses (the middle panel).
There, $P_3$ has a value ranging from $0.97 P$ to $1.03 P$, which results in the minimum repetition time $P^{\rm obs}_3\sim 33$ (see the bottom panel in Figure \ref{fig:obs_model_modeledpulses}). 
Note that the maximum value of the observed repetition time in the figure is clipped at $P^{\rm obs}_3 \sim 150 P$ as $P^{\rm obs}_3 \rightarrow \infty$ for $P_3 = 1 P$ when the apparent drift direction changes.
The figure clearly shows that we can model the observed non-stable drift characteristic in J1750$-$3503 with small variations in the actual repetition time when $P_3 \sim 1 P$. 

\section{Discussion} \label{sec:discussion}

Like the radio emission mechanism, the nature of the drifting phenomenon is a long-standing problem in pulsar astronomy.
Although the model for subpulse drift was proposed in the early stages of pulsar research, its main assumption, the motion of sparks around the magnetic axis, is questioned to this day \citep[see, e.g.][]{2016_Basu, 2020_Mitra}.
In  recent years,  progress  in both observational and theoretical studies has put us on the verge of fully understanding the phenomenon.
The increasing number of observed drifting pulsars has allowed determination of the dependence of the repetition time of the drift pattern $P_3$ on the rotational energy-loss rate $\dot{E}$ \citep{2016_Basu}.
Furthermore,  theoretical grounds of the LBC (Lagging Behind Corotation) model have been developed \citep[see the Appendix in][]{2020_Mitra}.

On the other hand, the notion that the drift velocity depends not on the absolute value, but on the variation of accelerating potential across the polar cap \citep{2012_Leeuwen} allowed development of the MC (Modified Carousel) model of drifting subpulses \citep{2017_Szary}.
In the MC model sparks rotate around the electric potential extremum at the polar cap.
If this extremum coincides with the center of the polar cap, the subpulses drift around the magnetic axis.
Both models can explain a variety of drifting data, when combined  with a non-dipolar surface magnetic field \citep[see, e.g.][]{2020_Basu, 2020_Szary}.
Although for each model the theoretical grounds have been laid out (see the main text of \citealt{2020_Basu} and the Appendix of \citealt{2020_Szary}),
they both currently lack the description of the mechanism that ensures the pattern of discharging regions is also stable. 
The detailed modeling behind this is more involved, and we are working on addressing it, in a future paper (Szary et al. 2023, \emph{in prep.}).

The first pulsar for which changes of drift direction were reported is \mbox{B0826$-$34} \citep{1985_Biggs, 2005_Esamdin}.
It is characterized with similar rotational properties to \J, with the characteristic age of $29\,{\rm Myr}$ and low rate of rotational energy loss $\dot{E}=6.2 \times 10^{30}\,{\rm erg\,s^{-1}}$ which also puts it close to the graveyard region in the $P-\dot{P}$ diagram.
There are two main differences between these pulsars, however.
First, \mbox{B0826$-$34} is effectively an aligned rotator with \mbox{$\alpha \approx 0.5^{\circ}$} and observed emission covers the whole rotational phase \citep[see, e.g., ][]{2005_Esamdin},
while \J~ is a slightly more inclined rotator with the inclination angle \mbox{$5^\circ \lesssim \alpha \lesssim 25^\circ$} (see Section \ref{sec:geometry}), implying that the rotation axis does not pass through the polar cap. 
Second, \mbox{B0826$-$34} appears to exhibit a weak mode, barely distinguishable from pulse nulling, for at least 70 per cent of the time. \J, on the other hand, seems not to exhibit any mode changing or clear nulling (see Section \ref{sec:nulling}).

The second pulsar for which changes of drift direction were reported is \mbox{B0540+23} \citep{1991_Nowakowski}. 
It is characterized with considerably different rotational properties to both \J~ and \mbox{B0826$-$34}, with period $P=0.246$ and period derivative \mbox{$\dot{P}=1.5\times 10^{-14} \,{\rm s/s }$}, which result in the characteristic age of $0.253\,{\rm Myr}$ and relatively high rate of rotational energy loss $\dot{E}=4.1 \times 10^{34}\,{\rm erg\,s^{-1}}$.
Similar to \J~ it is non-aligned rotator with $\alpha=38^{\circ}$ \citep{1993_Rankin} and relatively narrow pulse width, \mbox{$W_{10}=28.5^{\circ}$} \citep{1996_Gil}.
Likewise, it does not exhibit any nulls, however it is characterized with morphological changes that resemble the mode switching on time scales as short as five pulses \citep[see][]{1991_Nowakowski}.
Furthermore, for most of the time its subpulse drift is irregular without specified drift direction, and as a result there is no clear signature of $P_3$ in the fluctuation spectrum. 

Since in the aligned rotator case there is no difference in the predictions of the LBC and MC models, as in both cases subpulses would drift around the magnetic axis, in contrast to \mbox{B0826$-$34}, the observations of \J~ allow us to put some additional constrains on the drifting model.
Furthermore, the lack of obvious mode switching and its regular drift makes it an ideal candidate for studying the drifting phenomenon.
Below we discuss if and how the drifting properties of \J~ can help distinguish between the models.

\subsection{Intrinsic drift direction change}
\label{sec:discussion.direction}

In this section we discuss what is required to change the actual direction of plasma drift with respect to the neutron star rotation.

In the MC model the drift direction depends on the electric potential variation at the polar cap.
The spark structure is somehow stable, i.e. the distance between sparks does not change significantly while sparks rotate around electric potential extremum.
The sparks location and rotation is forced by the drift of plasma between sparks (see Figure 8 in \cite{2017_Szary}).
Note that this idea differs from the idea presented in \cite{2012_Leeuwen} where sparks (plasma columns) are surrounded by vacuum. 
However, there must be a mechanism which prevents or limits discharges in regions between sparks so it does not result in radio emission in those regions. 
One of the possible scenarios is a reverse plasma flow, induced by a mismatch between the magnetospheric current distribution and the current injected in the spark forming regions \citep[see, e.g.,][]{2012_Lyubarsky}.

There are two main factors which determine the drift direction in the MC model.
The first factor is the pulsar geometry and the location of the line of sight with respect to the electric potential extremum within the polar cap.
In general, a line of sight passing in between the rotational axis and the electric potential extremum, produces a drift direction opposite to that when both the rotational axis and the electric potential extremum are at the same side of the line of sight.
Thus, the drift direction can be modified by a change in location of the polar cap or a change in location of the electric potential extremum at the polar cap.
A change in polar cap location is possible by a change in structure of surface magnetic field while a change in electric potential across the polar cap may be caused by variation in plasma flow from the polar cap region.
In the MC model, for an inclined pulsar geometry, a relatively small shift of the order $(\beta/\rho) r_{\rm pc}$ may cause drift to change its direction, where $r_{\rm pc}$ is the polar cap radius. 
In the case of \J, assuming purely dipolar configuration of surface magnetic field, this requires a shift of the order of $\sim 100 \; \rm m$.
Note that if the surface magnetic field is highly non-dipolar a very small shift of the order of several meters may cause drift to change its direction.
However, for the aligned rotator, such as \mbox{B0826$-$34}, some other mechanism would have to exist.

The second factor that can influence the drift direction is the type of electric potential extremum at the polar cap.
In general, for a pulsar geometry ($\mathbf{\Omega} \cdot \mathbf{B}<0$) with a net positive charge at the polar cap, the spark forming regions, as plasma starved regions, result in electric potential minimum at the polar cap center.
On the other hand, an anti-pulsar geometry ($\mathbf{\Omega} \cdot \mathbf{B}>0$), with a net negative charge at the polar cap, is associated with electric potential maximum.
A change in the drift direction requires a significant change in electric potential variation, e.g. switching between a minimum and a maximum  at the polar cap center, which in principle could be caused by changing the reverse plasma flow in regions between sparks.
However, such drastic changes in the electric potential variation on the polar cap would most likely be associated with changes in the sparks structure, which is not visible in stable pattern of single-pulse emission in \J~ (see Figure \ref{fig:single_pulses}).

In the LBC model the drift direction is defined by the location of the polar cap with respect to the rotation axis.
By introducing a non-dipolar surface magnetic field, one can change the size, shape and location of the polar cap, and thus influence the observed drift characteristics \citep[see, e.g., ][]{2020_Basu}.
To change the drift direction, the position of the polar cap with respect to the rotation axis has to change.
Thus, it requires a change in configuration of surface magnetic field.
For an aligned rotator, a change in the polar cap position has to be of the order of at least the polar cap size, which for the non-dipolar configuration of surface magnetic field may range from a dozen to several dozen meters \citep[see, e.g.,][and references therein]{2017_Szary_XMM}.
For an inclined rotator, the magnitude of change depends on the initial configuration of surface magnetic field.
If initially an actual polar cap is near the magnetic axis, a change in the polar cap position has to be of the order of at least $(\alpha / 360^{\circ}) R$.  
Here $\alpha$ is the inclination angle in degrees and $R$ is the neutron star radius.
In the case of \J~ this requires a change in the polar cap position of the order of $\sim 300-600$ meters.
However, if the initial position of the polar cap is near the rotation axis, then a change in position has to be of the order of the polar cap size, as in an aligned rotator case.

In conclusion, in order to explain the intrinsic drift direction change both the MC and LBC models require very rapid significant changes either in electric potential variation across the polar cap or in the position of the polar cap itself. 

\subsection{Apparent drift direction change}
\label{sec:discussion.direction2}

In Section \ref{sec:aliasing} we have shown that the apparent drift direction can be interpreted as an aliasing phenomenon, and that we can reproduce the drifting behavior of \J~ with $P_3\sim P$ (or less for higher alias order) and small variations of the order of $6\%$.
Both the LBC and MC models have different predictions about the true value of $P_3$.

In the LBC model, using the simple assumption that lagging behind corotation is associated with the increasing subpulse phase for successive pulses, the anti-correlation \mbox{$P_3 =  (\dot{E} / \dot{E_0})^{-0.6 \pm 0.1}$}, was found, where \mbox{$E_0 = (2.3 \pm 0.2) \times 10^{32}\,{\rm erg\,s^{-1}}$} \citep{2016_Basu, 2019_Basu}.
That anti-correlation predicts the repetition time of \J~ to be $P_3=(10^{+6}_{-3}) P$.
This value seems to be inconsistent with the measured value \mbox{$P_3^{\rm obs}=(43.5 \pm 0.4) P$}.
It is worth mentioning, that if we consider only pulsars with positive drift (according to the definition of drift direction used in this paper) no such anti-correlation can be inferred (see, e.g., the blue circles in the right panel of Figure 1 in \cite{2019_Basu}).
This suggest that some other factors, for instance the non-dipolar structure of surface magnetic field, may play an important role in determining the actual direction of drift, and that further studies are required to verify the suggested correlation.
Note that the definition of positive drift in this paper is opposite to the definition used in \cite{2016_Basu, 2019_Basu}, and that there are some errors, for instance in caption of Figure 6 in \cite{2016_Basu}, related to the drift direction definition.

In the original carousel model, on the other hand, the dependence \mbox{$P_3 \simeq 5.6 \, n_{\rm sp}^{-1} \, B_{12} \,  P^{-2}$} was postulated, where $n_{\rm sp}$ is the number of sparks, and $B_{12}$ is the magnetic field in units $10^{12} \, {\rm G}$ \citep[see Equations 33 and 34 in][]{1975_Ruderman}.
Assuming a dipolar magnetic field at the surface, \mbox{$B_{\rm d}=3.2\times 10^{19} (P \dot{P})^{0.5}\, G$} and the spin down energy loss \mbox{$\dot{E}=4 \pi^2 I P^{-3} \dot{P}$}, where \mbox{$I\simeq 10^{45} \, {\rm g \, cm^2}$} is the moment of inertia, we can find the dependence $P_3 \simeq ({5.6}/{n_{\rm sp}}) \left ( {\dot{E}}/{\dot{E}_1} \right )^{0.5}$, where \mbox{$E_1 \simeq 4\times 10^{31} \;{\rm erg\,s^{-1}}$}.
Assuming thirteen sparks (see Section \ref{sec:aliasing}), this results in the repetition time $P_3 \simeq 0.15 P$, which is consistent with observations assuming alias order $n=7$.
This high alias order corresponds to the maximum possible drift velocity, i.e. the minimum possible value of $P_3$, and may be unlikely in reality.
In the MC model, for instance, the drift velocity depends not on the absolute value, as proposed by \cite{1975_Ruderman}, but on the variation of the accelerating potential across the polar cap.
Thus, in the MC model it is not possible to determine the dependency of $P_3$ on $\dot{E}$ without knowing the details of sparks formation and existence of regions where they do not arise.

In the case of B0826$-$34, \cite{2004_Gupta} constructed a model that explains its behavior assuming that the actual drift is much faster and we observed only its aliased value.
This claim was later called into question by \cite{2012_Leeuwen} who argued that since the observed reversals influence the intrinsic subpulse brightness the reversals are not just apparent.
Furthermore, detailed analysis on a number of other pulsars also concluded their pulse trains were unaliased  \citep[see, e.g.,][]{lsr+03}. 
However, in the MC model, a change in $P_3$ should be connected with variation in accelerating potential at the polar cap and in principle may have some implications on subpulse brightness.
Therefore, we should not exclude the possibility of aliasing in B0826$-$34 based only on intrinsic subpulse brightness changes. 

In the case of \J~ there is no evidence of an intrinsic change of subpulse brightness during reversals (see Section \ref{sec:average}).
However, the average profile characteristics and the separation between subpulses change when drift changes direction (see Section \ref{sec:p2_single-pulse}).
As pointed out by \cite{2004_Gupta}, the measured value of $P^{\rm m}_2$ is a ``Doppler-shifted'' version of the actual subpulse separation, $P^{\rm t}_2$, and it depends on the drift rate: $P^{\rm t}_2 = P^{\rm m}_2 (1 - D^{\rm t} / 360^{\circ})$, here $D^{\rm t}$ is the true value of drift rate.
Thus, a varying nature of drift rate is naturally associated with a change in average profile.
The drift rate change inferred from the subpulse separation ($\sim 7\%$) is consistent with a drift rate change required to explain the observed behavior of \J~by aliasing.   

In the recent work of \cite{2022_Janagal} it is proposed that the four observed drift modes in J1822$-$2256 are the result of intrinsic $P_3 \sim P$ (or less for higher alias order) and a varying number of sparks.
It is worth pointing out that although all the drift modes in J1822$-$2256 have the same direction in principle different spark configurations (or varying drift rates) in combination with $P_3 < 2P$ could result in opposite drift directions in different modes.

\section{Conclusions} \label{sec:conclusions}
In this paper we present the unusual subpulse modulation in \J. 
Although we cannot exclude that the changes in drift direction are intrinsic, and thus cannot rule out the validity of the 
Lagging Behind Corotation model, we show that the observed data is in full agreement with the Modified Carousel model. 
The  behaviour in \J~  then self-consistently follows from an actual repetition time $P_3 \sim P$, where we observe its aliased value. 

\vspace*{0.5cm}

\acknowledgments

The MeerKAT telescope is operated by the South African Radio Astronomy Observatory, which is a facility of the National Research Foundation, an agency of the Department of Science and Innovation.
This research received funding from the Netherlands Organisation for Scientific Research (NWO) under project ``CleanMachine'' (614.001.301)
 and through Vici research programme `ARGO' (639.043.815); 
and from the European Research Council under the European Union’s Seventh Framework Programme (FP/2007-2013) / ERC Grant Agreement n. 617199.
GW thanks the University of Manchester for Visitor status.





\bibliographystyle{aasjournal}

\end{document}